\documentclass[twocolumn,showpacs,preprintnumbers,amsmath,amssymb]{revtex4}
\date{\today}

\usepackage{graphicx}

\newcommand{\rr}{\mbox{\boldmath $r$}}
\newcommand{\EE}{\mbox{\boldmath $E$}}

\newcommand{\hh}{\mbox{\boldmath $h$}}

\newcommand{\OOmega}{\mbox{\boldmath $\Omega$}}
\begin{document}

%%
%% Title page
%%

\title{
Wave Chaos in Rotating Optical Cavities
}

\author{Takahisa Harayama, Satoshi Sunada, and Tomohiro Miyasaka}
\affiliation{
Department of Nonlinear Science ATR Wave Engineering Laboratories
2-2-2 Hikaridai Seika-cho Soraku-gun Kyoto 619-0228 Japan\\  
}

\begin{abstract}
%%%%%%%%%%%%%%%%%%%%%%%%%%%%%%%%%%%%%%
%
%  ABSTRACT
%
%%%%%%%%%%%%%%%%%%%%%%%%%%%%%%%%%%%%%%
%We study theoretically and numerically 
%the effect of rotation on two types of eigenmodes; modes localizing on 
%stable periodic trajectories and wave-chaotic modes, which do not
% localize on any ray-trajectories, in optical
% cavities that exhibit the ray-dynamical properties of a mixed system.       
%The frequency splitting due to rotation on the former modes 
%of non-rotating cavity can be explained as the Sagnac effect. 
%We show that, even when the eigenmodes are wave-chaotic,
It is shown that, even when the eigenmodes of an optical cavity are wave-chaotic,
the frequency splitting due to the rotation of the cavity occurs and the frequency
 difference is proportional to the angular velocity although the splitting
 eigenmodes are still wave-chaotic and do not correspond to any
 unidirectionally-rotating waves.
\end{abstract}


\maketitle

%%
%% Start of the main text
%%

%%
%% 1. Introduction
%%

%% 1-1. Review of Sagnac Effect

In 1913 Sagnac pointed out that the path length of 
the clockwise (CW) propagating light 
in a rotating ring interferometer for one round trip  
is different from that of the counter-clockwise 
(CCW) propagating light and hence the Sagnac interferometer 
can be used as a rotation rate sensor \cite{Post, Crow}. 

%The difference of the path lengths between counter-propagating 
%lights had been experimentally observed as the phase difference
%for about 50 years after the finding of the Sagnac effect.  
%However, the Sagnac interferometer could not take the place of 
%the mechanical gyroscopes although nowadays the measurement of 
%the phase difference in the Sagnac interferometer has become very precise 
%due to the technology of optical fibers.  
%The development of the laser in 1960's changed the situation completely. 
%In the case of a rotating ring laser, 
%the difference of the path lengths between counter-propagating 
%lights is observed very precisely as the frequency difference 
%between the CW and CCW lasing modes instead of the phase difference. 
%Ring laser gyroscopes and fiber optic gyroscopes 
%are now used in airplanes, rockets, and ships etc. 
%for autonomous navigation \cite{Crow, }. 

The conventional theory of ring laser gyroscopes is based on 
the Sagnac's original idea, that is, the difference of 
the path lengths between counter-propagating lights  
derived from the special theory of relativity 
for ray-dynamical description \cite{Post, Crow}. 
More precise electromagnetic wave description of 
the Maxwell equations including the effect of rotation is given by  
the general theory of relativity \cite{Post, Crow, Landau}.  
These two different conventional ways describing the effect of rotation  
on the lasing frequency has given the same result 
because of the following two assumptions:
(i)The ray-wave correspondence of the cavity-modes; 
the conventional theory assumes that there exist closed ring trajectories 
and the cavity-modes localize on these ring trajectories in a ring resonator. 
(ii)The equivalence of the descriptions for the cavity modes by 
standing waves and propagating waves; 
the conventional theory assumes that the size of the ring resonator 
is much larger than the wavelength of the laser, and 
in this short wavelength limit the cavity-modes are separated into 
two directions propagating along and transversal to the ring trajectory. 
 
The assumption (ii) means that the cavity-modes can be always expressed 
as the unidirectionally rotating waves and the counter-propagating modes 
are degenerate. Precisely speaking, however, the cavity-modes 
of non-rotating cavities are the standing waves 
as far as the shape of the cavity does not have 
a special symmetry like a circle. 

%We have recently established the wave-dynamical approach to 
%the rotating resonant cavity without assumption (i) and (ii), 
%and shown that the nearly degenerate standing wave cavity-modes of 
%the non-rotating cavity which localize on the ring trajectory 
%and hence satisfy the assumption (i) 
%change into the pair of the counter-propagating waves 
%when the angular velocity is larger 
%than a certain threshold \cite{SH1,SH2}. 
%The frequency difference start to increase 
%above this threshold and become proportional to the angular velocity.  
%The existence of the threshold for the transition from the standing wave 
%to the rotating wave never be found by the conventional theory 
%because of the assumption (ii). 

%In this Letter, we discuss the case which does not even satisfy 
%the assumption (i) as well as (ii). 
The researches on quantum chaos for these three decades have shown 
that there exist so-called wave-chaotic cavity-modes 
that do not localize on any ray-dynamical trajectories 
due to the chaotic property of the ray-dynamics in the cavity
\cite{Stone, Gutzwiller, Stockmann}.
The ray-dynamical description for ring laser gyroscopes 
is not applicable for these wave-chaotic 
cavity-modes. On the other hand, the wave-dynamical description 
is still applicable. 
%% 1-3. Main Claim
%%% 1-3-1. Sagnac Effect in Resonant Microcavities
In this Letter, 
the recently established theory \cite{SH1,SH2} of rotating resonant microcavities 
without the assumptions (i) and (ii)  
is applied to the cavity which shows the ray-dynamical properties 
of the mixed system. It is well-known that some eigenfunctions 
in the mixed system localize on the stable periodic trajectories 
while others are wave-chaotic and spread over the chaotic sea 
in the phase space. 
We show that the degenerate eigen-frequency corresponding to 
the wave-chaotic cavity-mode of the non-rotating cavity 
splits into two frequencies and their difference is 
proportional to the rotation rate although the splitting cavity-modes 
are still wave-chaotic and do not have any corresponding CW and CCW 
propagating modes as well as ray-dynamical counterparts, 
which cannot be explained by the conventional Sagnac effect.

%%
%% End of Introduction
%%

First let us introduce the ray-dynamical properties 
and the wavefunctions of the eigenmodes 
when the cavity is not rotating. 
%%%%%%
%%
%% 2. System
%%
%%%%%%
%%
%% 2-1. Shape of the cavity
%%
%%%%%
We will discuss the two-dimensional shape of the resonant microcavity 
defined by $R(\theta)=R_0(1+\epsilon \cos 4\theta)$ 
in the cylindrical coordinates.
The parameters are set as follows:
$R_0 = 6.2866\mu m, \epsilon = 0.04$,
and the refractive index $n = 1$.
%%%%% 
%%
%% 2-2. Ray-dynamics
%%
%%%%%

\begin{figure}
\begin{center}
%\hspace{10mm} 
%\raisebox{0.0cm}{\includegraphics[height=5cm,width=4cm,angle=270]{map.eps}} 
\raisebox{0.0cm}{\includegraphics[width=7cm,height=5cm]{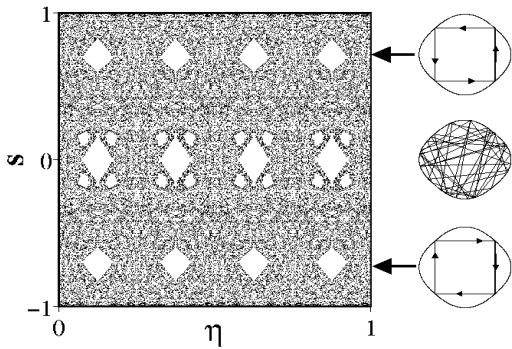}}
\end{center}
\vspace{-4mm}
\caption{\label{fig1} 
The Poincar\'{e} surface of the section on the Birkhoff coordinate.}
\end{figure}

Ray-dynamical properties can be easily seen by 
plotting the trajectories on the Poincar\'{e} 
surface of the section (SOS). 
It is convenient to use the Birkhoff coordinate $(\eta,s)$ for SOS, 
where $\eta$ is the normalized curvilinear distance measured along 
the edge of the cavity from a certain origin on the edge 
to the incident point and $s$ is the sine of the incident angle. 
The SOS of ray-dynamics in this cavity shown in Fig.~\ref{fig1} 
includes stable islands and a chaotic sea, which shows 
it is a typical mixed system. 
The stable islands around $s=  +(-)1/\sqrt{2}$ correspond to CCW(CW) rotational
ray-motions along a ring trajectory,
as the ray-motion shown at the top (bottom) of the right hand side in
Fig. \ref{fig1}.
Then, typical example of a single trajectory in the chaotic sea are shown at
the middle.

%%%%%
%%
%% 2-3. Wave functions (no rotation rate)
%%
%%%%

\begin{figure}
\begin{center}
%  \begin{tabular}{ c c }
%\hspace{10mm} 
%\raisebox{0.0cm}{\includegraphics[width=3.4cm]{wfr0-odd.eps}}
%    &
%    \includegraphics[width=3.4cm]{wfr0-even.eps}\\
%    (a) & (b) \\
%\raisebox{0.0cm}{\includegraphics[width=3.8cm]{ht-odd.eps}}
%    &
%    \includegraphics[width=3.8cm]{ht-even.eps}\\
%    (c) & (d) 
%\hspace{10mm} 
%  \end{tabular}
\raisebox{0.0cm}{\includegraphics[width=8cm]{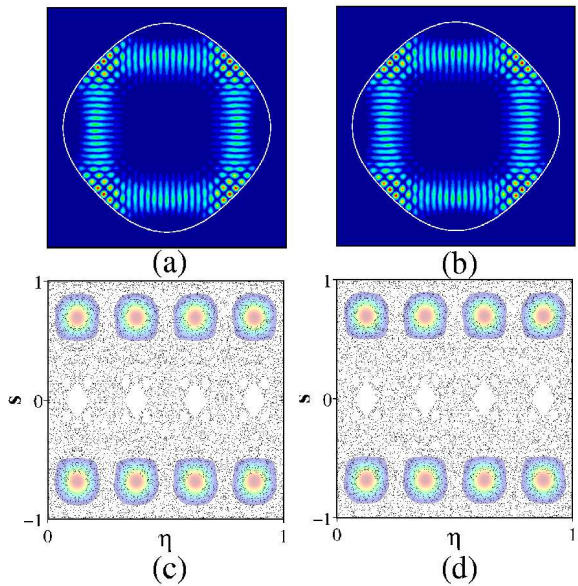}}
\end{center}
\vspace{-4mm}
\caption{\label{fig2} 
(Color)
(a-b) The eigenmodes corresponding to the stable islands
 of the ring trajectory in non-rotating cavity. 
Each eigenmodes labeled by (a) and (b) are odd(even) parity and even(odd) parity with
 respect to the $x(y)-$ axis, and they are obtained at $nkR_0= 50.264118$. 
(c-d) The Husimi representations corresponding to the eigenmodes (a) and
 (b). 
}
\end{figure}

\begin{figure}
\begin{center}
%  \begin{tabular}{ c c }
%\hspace{10mm} 
%\raisebox{0.0cm}{\includegraphics[width=3.4cm]{wf0-odd.eps}}
%    &
%    \includegraphics[width=3.4cm]{wf0-even.eps}\\
%    (a) & (b) \\%
%
%\raisebox{0.0cm}{\includegraphics[width=3.8cm]{hc-odd.eps}}
%    &
%    \includegraphics[width=3.8cm]{hc-even.eps}\\
%    (c) & (d) 
%\hspace{10mm} 
%  \end{tabular}
\raisebox{0.0cm}{\includegraphics[width=8cm]{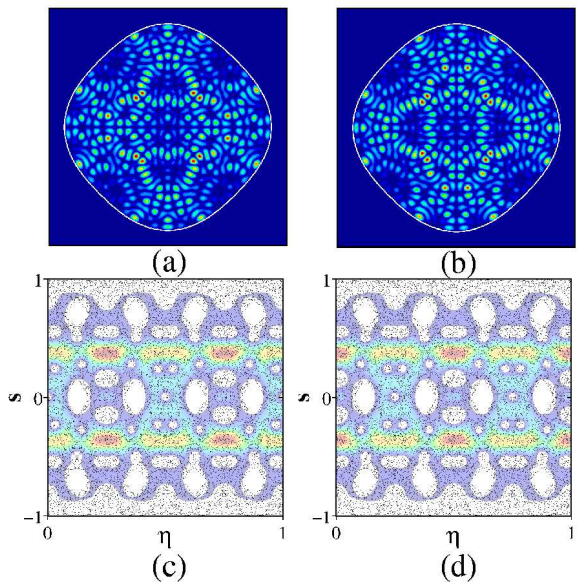}}
\end{center}
\vspace{-4mm}
\caption{\label{fig3}
(Color)
(a-b) The wave-chaotic eigenmodes in non-rotating cavity. 
Each eigenmodes labeled by (a) and (b) are odd(even) parity and even(odd) parity with
 respect to the $x(y)-$ axis, and they are obtained at $nkR_0=
 50.264118$, which is the almost same value as that of the modes shown in Fig. \ref{fig2}.
(c-d) The Husimi representations corresponding to the eigenmodes (a) and
 (b). 
}
\end{figure}

The wave properties are discussed only in the case of 
TM mode of electromagnetic field oscillating as 
$\EE(\rr,t) = (0,0,\psi(\rr)e^{-ick t}+c.c.)$ where $c$ is 
the velocity of light and the eigenfrequency $\omega$ equals $ck$, and   
the Dirichlet boundary condition are imposed for simplicity.
By solving the Helmholtz equation 
$
\left(
{\nabla_{xy}}^2 + n^2 k^2
\right)\psi
= 0  
$ 
where $n$ is the refractive index inside the cavity,  
we obtain the eigenmodes respectively corresponding to 
the stable islands and the chaotic sea as shown in Fig.~\ref{fig2} (a)
and \ref{fig3} (a).
%Fig. \ref{fig2} (c) and \ref{fig3} (c) 
%show the Husimi representations \cite{Crespi} of the eigenmodes
%shown in Fig. \ref{fig2} (a) and \ref{fig3} (a), respectively.
The eigenmode in the Husimi representation \cite{Crespi} of Fig. \ref{fig2} (c)
localizes on both the stable islands of the ring trajectories at
$s=\pm 1/\sqrt{2}$, which means that the eigenmode in
the non-rotating cavity has both CW and CCW
propagating wave components and is standing wave along the ring
trajectory.
On the other hand, Fig. \ref{fig3} (c) shows that the eigenmode in
Fig. \ref{fig3} (a) does not localize on the stable islands but
is distributed over the chaotic sea. As well known, 
no simple ray-wave correspondence exists for such an eigenmode. 

% standing waves 
%From the Husimi representations shown in Fig. 2, 
%one can see that these eigenmodes have CW and CCW-propagating wave
%components and are standing waves. 
%  

These two modes belong to the same symmetry class 
that is respectively even and odd 
with respect to the $y$-axis and the $x$-axis 
because the cavity has $C_{4v}$ symmetry. 
When these eigenmode are rotated by $\pi/2$, one can obtain 
the other eigenmodes that are respectively even and odd 
with respect to the $x$-axis and the $y$-axis.  
These modes created by $\pi/2$ rotation are orthogonal to 
the original ones. 
Thus, the pairs of degenerate eigenmodes in this cavity are obtained 
as shown in Fig.~\ref{fig2} (b) and \ref{fig3} (b), 
and the Husimi representations of the
eigenmodes shown in Fig. \ref{fig2} (b) and \ref{fig3} 
(b) are shown in Fig. \ref{fig2} (d) and \ref{fig3} (d), respectively. 

\begin{figure}
\begin{center}
%  \begin{tabular}{ c c }
%\hspace{10mm} 
%\hspace{10mm} 
%\raisebox{0.0cm}{\includegraphics[width=3.4cm]{wfr1.eps}}
%    &
%    \includegraphics[width=3.4cm]{wfr1.eps}\\
%    (a) & (b) \\
%\raisebox{0.0cm}{\includegraphics[width=3.8cm]{htr1.eps}}
%    &
%    \includegraphics[width=3.8cm]{htr2.eps}\\
%    (c) & (d) \\
%\raisebox{0.0cm}{\includegraphics[width=3.0cm,angle=270]{at1.ps}}
%    &
%    \includegraphics[width=3.0cm,angle=270]{at2.ps}\\
%    (e) & (f) 
%  \end{tabular}
\raisebox{0.0cm}{\includegraphics[width=8.7cm]{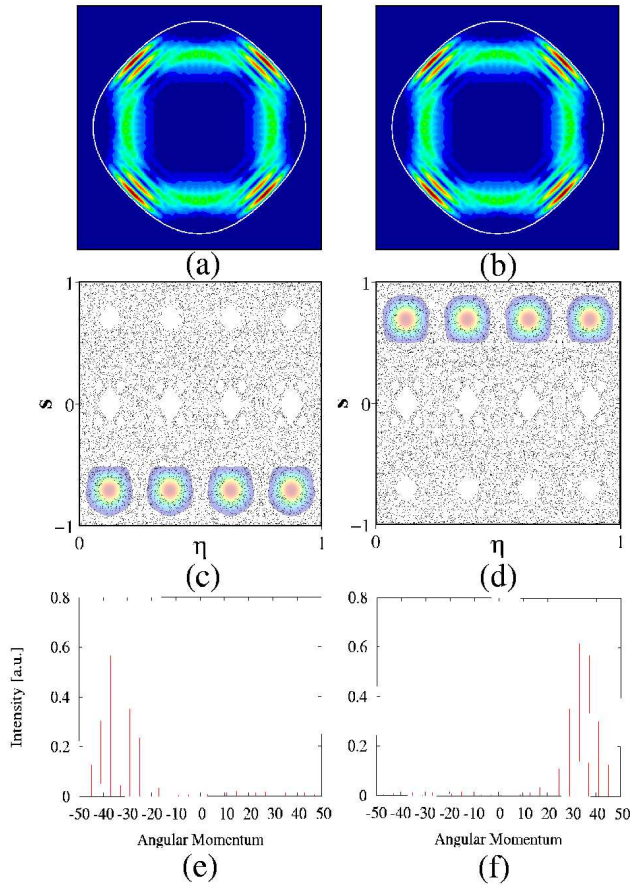}}
\end{center}
\vspace{-4mm}
\caption{\label{fig4}
(Color)  
(a-b) The eigenmodes corresponding to the stable islands
 of the ring trajectory in rotating cavity with $R_0\Omega/c=6.28 \times 10^{-5}$. 
(c-d) the Husimi representations corresponding to the eigenmodes (a) and
 (b). (e-f) the angular momentum spectrum of the eigenmodes (a) and (b).
}
\end{figure}

\begin{figure}
\begin{center}
%  \begin{tabular}{ c c }
%\hspace{10mm} 
%\raisebox{0.0cm}{\includegraphics[width=3.4cm]{wf1.eps}}
%    &
%    \includegraphics[width=3.4cm]{wf1.eps}\\
%    (a) & (b) \\
%\raisebox{0.0cm}{\includegraphics[width=3.8cm]{hcr1.eps}}
%    &
%    \includegraphics[width=3.8cm]{hcr2.eps}\\
%    (c) & (d) \\
%\raisebox{0.0cm}{\includegraphics[width=3.0cm,angle=270]{ac1.ps}}
%    &
%    \includegraphics[width=3.0cm,angle=270]{ac2.ps}\\
%    (e) & (f) 
%\hspace{10mm} 
%
%  \end{tabular}
\raisebox{0.0cm}{\includegraphics[width=8.7cm]{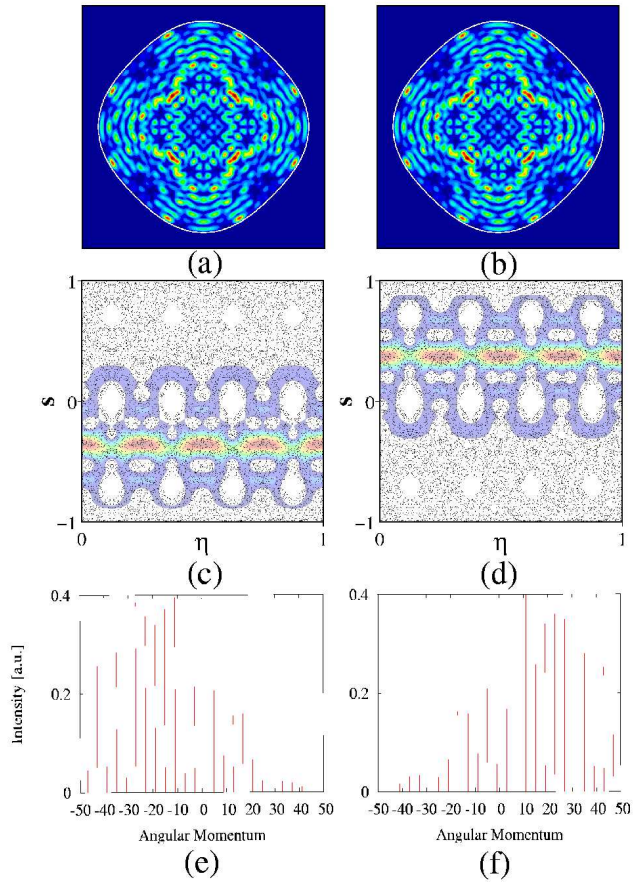}}
\end{center}
\vspace{-4mm}
\caption{\label{fig5} 
(Color) 
(a-b) The wave-chaotic eigenmodes in rotating cavity with 
$R_0\Omega/c=6.28 \times 10^{-5}$. 
(c-d) the Husimi representations corresponding to the eigenmodes (a) and
 (b). (e-f) the angular momentum spectrum of the eigenmodes (a) and (b).
}
\end{figure}

%%%%%
%%
%% Review of the new theory  
%%
%%%%%

Next we discuss the effect of rotation of the cavity on 
the above eigenmodes. 
%In the conventional theory of ring laser gyroscopes, 
%the frequency shift of the eigenmodes proportional to the angular 
%velocity of the rotating ring cavity is derived 
%by assuming that the electric field propagates 
%one-dimensionally along the ring trajectories \cite{Crow,}.  
%This method is applicable only to the eigenmodes 
%where the electric fields localize on the stable islands.  
%The more general theory of the effect of the rotating resonant
%microcavity 
The theory of the effect of the rotating resonant microcavity 
on the resonances has recently been obtained \cite{SH1,SH2} and is applicable 
to the wave-chaotic eigenmodes where the electric fields spread over 
the whole phase space. 
Let us briefly review this theory for the case of the degenerate
eigenmodes.  

%%
%% 2. Maxwell Equations for Rotating Microcavities
%%

%% 2-1. Maxwell equations in an arbitrary gravitational field

According to the general theory of relativity,  
the electromagnetic fields in a rotating resonant microcavity 
are subject to the Maxwell equations generalized to a non-inertial 
frame of reference in uniform rotation with angular velocity vector 
$\OOmega$ \cite{Post,Crow,Landau}. 
By neglecting $O(|\OOmega|^2)$  
%% 2-4. Equations for Stationary States 
and assuming the 2D resonant microcavity is  
perpendicular to angular velocity vector $\OOmega$, 
we obtain the following equation for the eigenmodes 
of the rotating cavity,  

%%% (2) %%%
\begin{eqnarray}
\left(
{\nabla_{xy}}^2 + n^2 k^2
\right)\psi
-
2ik
\left(\hh\cdot\nabla
\right)\psi
= 0, 
\label{fundeq}
\end{eqnarray}
where $\hh = (\rr \times \OOmega)/c$ and 
the 2D resonant cavity is rotating on $xy$-plane clockwisely,i.e.,  
$\OOmega=[0,0,\Omega (>0)]$. 

%%
%% 3. Theory of Transition from Standing Waves to Rotational Waves 
%%

By using the perturbation theory for degenerate states 
in quantum mechanics, the degenerate eigenmodes $\psi_0$ and $\psi_1$
of the non-rotating cavity are superposed to reproduce 
the solutions of Eq.~(\ref{fundeq}) as 
\begin{eqnarray}
\psi_{\pm} = 1/\sqrt{2}\psi_0 \pm i/\sqrt{2}\psi_1,
\label{expan2}
\end{eqnarray} 
where the degenerate wave number $k_0$ splits into two. 
%up to the first order of $|\Omega/c|$, 
%\begin{eqnarray}
%k =k_0 \pm \dfrac{1}{n^2}\left|\int\int_{D} d\rr
%\psi_0 \left(\hh\cdot\nabla\right)\psi_1 \right|. \label{eq:k-sag}
%\end{eqnarray}
Therefore, the frequency difference $\Delta \omega$ 
between the two eigenfunctions newly produced by rotation of the cavity 
is proportional to the angular velocity, 
\begin{equation}
\Delta \omega = 2
\left|\int\int_{D} d\rr
\psi_0 \left(y\frac{\partial}{\partial x}
-x\frac{\partial}{\partial y}\right)\psi_1 \right|
\frac{\Omega}{n^2}. \label{eq:w-sag}
\end{equation}

%%%
%
% Apply the new theory to the present cavity
%
%%%

Applying this theory to the degenerate eigenmodes where 
the electric fields localize on the ring trajectories 
as shown in Fig.~\ref{fig2}(a) and \ref{fig2}(b) yields almost the same results 
as the conventional theory for ring laser gyroscopes 
because the superposition (\ref{expan2}) 
of these degenerate eigenmodes of the non-rotating cavity 
produce the split eigenmodes of the rotating cavity 
which almost propagate CW and CCW along the ring trajectories. 
The eigenmodes become completely CW and CCW propagating modes 
in the short wavelength limit because it has been shown that 
the wave functions constructed by the Gaussian beams 
unidirectionally propagating along the ring trajectory are 
the eigenmodes of the cavity in the short wavelength limit \cite{Hakan}. 

However, when we apply this theory to the degenerate eigenmodes 
where the electric fields are wave-chaotic as shown in Fig.~\ref{fig3}(a) and \ref{fig3}(b),  
we cannot expect the superposed solutions (\ref{expan2}) 
are CW and CCW propagating modes. 
The cavity-mode $\psi_0$ in Fig.~\ref{fig3}(a) 
can be expressed by the superposition of the Bessel functions as
\begin{eqnarray}
\psi_0 & = & \sum_{m=0}^{\infty}  
a_{4m+1} J_{4m+1} (n k_0 r) \sin{(4m+1)\theta}  \nonumber \\ &+ & 
\sum_{m=0}^{\infty}
a_{4m+3} J_{4m+3} (n k_0 r) \sin{(4m+3)\theta}
\label{eqpsi0}, 
\end{eqnarray}
because the wave function $\psi_0$ is odd (even) 
with respect to $x$ $(y)$-axis. 
Then, the expression of the cavity-mode $\psi_1$ in Fig.~\ref{fig3}(b) can be 
obtained by $\pi/2$ rotation of $\psi_0$ , 
\begin{eqnarray}
\psi_1 
%& = & \sum_{m=0}^{\infty} 
%a_{4m+1} J_{4m+1} (n k_0 r) 
%\sin{(4m+1)\left(\theta-\frac{\pi}{2}\right)} \nonumber \\  
%& + & \sum_{m=0}^{\infty}
%a_{4m+3} J_{4m+3} (n k_0 r) 
%\sin{(4m+3)\left(\theta-\frac{\pi}{2}\right)} \nonumber\\ 
& = &  
\sum_{m=0}^{\infty} 
-a_{4m+1} J_{4m+1} (n k_0 r) \cos{(4m+1)\theta} \nonumber\\ 
& + & \sum_{m=0}^{\infty}
a_{4m+3} J_{4m+3} (n k_0 r) \cos{(4m+3)\theta}
\label{eqpsi1}, 
\end{eqnarray}
which has the same eigen-frequency as that of $\psi_0$ and  
is even (odd) with respect to $x$ $(y)$-axis. 
Accordingly, we have alternative expressions of the degenerate eigen-modes 
(\ref{expan2}),
\begin{eqnarray}
\psi_{\pm} 
& = & \mp \frac{i}{\sqrt 2}\sum_{m=0}^{\infty}a_{4m+1} J_{4m+1} (n k_0 r) 
\exp\left\{ \pm i (4m+1) \theta \right\} \nonumber \\ 
& \pm & \frac{i}{\sqrt 2} \sum_{m=0}^{\infty}a_{4m+3} J_{4m+3} (n k_0 r)
\exp\left\{ \mp i (4m+3) \theta \right\}. \nonumber \\ 
& & 
\label{eqpchaos} 
\end{eqnarray}
It is important that $\psi_\pm $ should be the eigen-mode of 
the rotating cavity, but it contains both CW and CCW  propagating waves. 
Consequently it is impossible to discuss the difference of 
the path lengths between CW and CCW propagating lights. 
Nevertheless, the frequency splitting  (\ref{eq:w-sag}) 
is still proportional to the angular velocity 
like the result from the conventional Sagnac effect. 
Therefore, it is concluded that the frequency splitting 
proportional to the rotation rate 
is the more general effect of rotation of the resonant cavity on 
its eigenmodes, and  only in the special case that the eigenmodes 
localize on the ray-dynamical ring trajectories, 
this frequency splitting can be related to the Sagnac effect that is 
the difference of the path lengths between the CW and CCW 
propagating modes.   

In order to confirm the above conclusion from the perturbation theory, 
we actually solved Eq.~(\ref{fundeq}) numerically in the rotating cavity
with angular velocity $\Omega$, and obtained the eigenmodes, as shown in
Fig. \ref{fig4} and Fig. \ref{fig5}. 
%and obtained the eigenmodes shown in Fig.~4.
%
%Then we obtained the eigenmodes corresponding to the stable island, as shown
%in Fig. \ref{fig4} (a) and (b), 
%and the wave-chaotic eigenmodes shown in Fig. \ref{fig5} (a) and (b)
%, respectively. 
%The Husimi representations of 
%the eigenmodes of Fig. \ref{fig4} (a) and (b) on the Birkhoff
%coordinates are shown in Fig. \ref{fig4} (c) and (d), respectively.
%On the other hand, 
%Fig. \ref{fig5} (c) and (d) show the Husimi representations of the
%wave-chaotic eigenmodes in Fig. \ref{fig5} (a) and (b),
%respectively.

\begin{figure}
\begin{center}
%\hspace{10mm} 
\raisebox{0.0cm}{\includegraphics[width=4cm,angle=270]{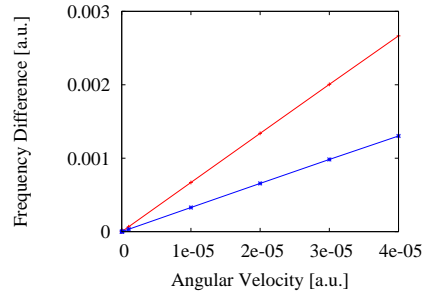}}
\end{center}
\vspace{-4mm}
\caption{\label{fig6} 
(Color)
The frequency difference 
between the eigenfrequencies of modes shown in Fig. \ref{fig4} versus the angular velocity
 (red).  The frequency difference between those of modes in
 Fig. \ref{fig5} versus the angular velocity (bule).
}
\end{figure}

%First, let us explain the effect of rotation on the eigenmodes
%corresponding to the stable island. 
Fig. \ref{fig4} (a) and (b) 
show the wavefunctions of the eigenmodes corresponding to the
stable island of the ring trajectory in rotating cavity.
Then the Husimi representations of the
eigenmodes of Fig. \ref{fig4} (a) and (b) are shown in Fig. \ref{fig4}
(c) and (d),
 and the distribution of the angular momentum components
$|a_m|^2$ of the eigenmodes are shown in Fig. \ref{fig4} (e) and (f),
respectively.
As seen in these figures, 
%the modes in these Husimi representations are distributed either in the
%stable island of CW rotating path along the ring trajectory or in the
%island of CCW rotating path, while the modes shown in Fig. \ref{fig2}
%(c) and (d) are distributed in both stable island when the cavity is not
%rotated.
the eigenmodes become almost unidirectionally propagating waves along the ring
trajectory in the rotating cavity, while they are the standing waves
along the ring trajectory in non-rotating cavity.
%
%The Husimi representations shown in Fig. 4 (c) and (d) show that the
%eigenmodes corresponding to the stable island change into the CW and CCW
% propagating wave along the ring trajectory, while they are standing
% waves for the angular velocity $\Omega = 0$, as shown in Fig. 2 (c) and (d). 
Then, the frequency splitting is proportional to the angular velocity,
as shown in Fig. \ref{fig6},
and the scale factor agrees with that calculated with the difference of
the path lengths of the ring trajectory.
That is, the conventional theory of the Sagnac effect 
can be reproduced in this case. 

Fig. \ref{fig5} (a) and (b) show the wave-chaotic eigenmodes, and 
the Husimi representations corresponding to each eigenmodes are shown in
Fig. \ref{fig5} (c) and (d). 
Unlike the case where the eigenmodes localize on the stable island, 
the distributions of the wave-chaotic modes in the Husimi
representations clearly show that 
the eigenmodes do not 
become unidirectionally-propagating modes corresponding to the stable 
islands. 
%spread not only in the chaotic sea of the $s < (>) 0$ region 
%but also in the chaotic sea of the $s > (<) 0$ region,
%although the distributions have disproportionate weight in one or other
%of two regions.
%This fact means that 
Moreover,
the angular momentum spectrum in Fig. \ref{fig5} (e) and (f) show 
the wave-chaotic eigenmodes have both the CW ($m<0$) and CCW ($m>0$)
propagating wave components even when the cavity is rotated.
However, the eigenfrequency actually splits into two, 
and the frequency splitting is proportional to the angular velocity 
as shown in Fig.~\ref{fig6}.  
Thus, one can confirm that the frequency splitting due to rotation can be
observed even on the eigenmodes which do not split into CW and CCW propagating
wave modes.

%\begin{acknowledgments}
The work was supported by 
the National Institute of information and Communication Technology
of Japan. 
%the Telecommunications Advancement Organization of Japan. 
%\end{acknowledgments}

%
%

\end{document}